\documentclass[onecolumn,showpacs,prl,superscriptaddress,preprint]{revtex4}
\usepackage{graphicx}
\begin{document}

\title{Electronic structure of epitaxial graphene nanoribbons on SiC $(0001)$}

\author{I. Deretzis}
\email{ioannis.deretzis@imm.cnr.it}
\affiliation{Scuola Superiore, Universit\`{a} di Catania, I-95123 Catania, Italy}
\affiliation{CNR-IMM, I-95121 Catania, Italy}

\author{A. La Magna}
\affiliation{CNR-IMM, I-95121 Catania, Italy}
\date{\today}

\begin{abstract}
We present electronic structure calculations of few-layer epitaxial graphene nanoribbons on SiC $(0001)$. Trough an atomistic description of the graphene layers and the substrate within the extended H\"{u}ckel Theory and real/momentum space projections we argue that the role of the heterostructure's interface becomes crucial for the conducting capacity of the studied systems. The key issue arising from this interaction is a Fermi level pinning effect introduced by dangling interface bonds. Such phenomenon is independent from the width of the considered nanostructures, compromising the importance of confinement in these systems.
\end{abstract}

\maketitle

It is nowadays commonly accepted that epitaxial graphene on silicon carbide substrates represents a viable method of controllable growth for the fabrication of high-quality graphene wafers \cite{2009NatMa...8..203E}. Indeed, recent experiments have focused on growth techniques on both the Si $(0001)$\cite{2009NatMa...8..203E,2008PhRvB..78x5403V} and the C $(000\bar{1})$ faces \cite{2008arXiv0812.4351C} of various SiC polytypes, whereas theoretical investigations have dealt with the composite substrate-graphene layer problem on two-dimensional structures considering few-atom supercells and exploiting lattice periodicity \cite{2007PhRvL..99g6802M,2008NanoL...8.4464P,2007PhRvL..99l6805V}. The universally accepted concept is that the process of graphitization takes place with the formation of an interface carbon layer (buffer layer) which decouples the electronic properties of the substrate from those of the subsequent graphene layers. The surface reconstruction is still a subject of debate for the $(000\bar{1})$ case, where $R2^{\pm}/R30^{\circ}$ rotational disorder and a high concentration of stacking faults has been been observed in the literature \cite{2008PhRvL.100l5504H}. On the other hand a $6\sqrt3\times6\sqrt3R30^{\circ}$ reconstruction has been verified for the (0001) surface \cite{2007PhRvB..76x5406R}, allowing the formation of both covalent and unsaturated bonds in the heterostructure's interface area. This study focuses on the Si-face grown films, where although the theoretical/experimental framework seems in agreement (e.g. $n$-type doping effects \cite{2009NatMa...8..203E,2007PhRvL..99g6802M}), there are still questions to be answered, mainly concerning the measured mobilities found to be lower ($2000 cm^{2}V^{-1}s^{-1}$ at low temperatures \cite{2009NatMa...8..203E}) than respective ones for films deposited on $SiO_2$, as well as the impact of one-dimensional (1D) confinement.

In this letter we theoretically study the electronic structure of 1D armchair graphene nanoribbons (AGNRs) with one or more  layers grown on SiC(0001), considering an atomistic description of the epitaxial layers and the substrate. The quantum chemistry is used within the extended H\"{u}ckel Theory on an $sp^3d^5$ Slater-type basis that considers both valence and polarization orbitals. This approach has been followed for the electronic structure study of 1D carbon allotrope structures \cite{2006JAP...100d3714K,2008PhRvB..77x5434R} and allows for the study of relatively large complexes (up to 4.4 nm wide in this work) respecting at the same time the chemical environment. The Slater parameters used here have been extracted from DFT calculations on SiC \cite{2000PhRvB..61.7965C} and have been extensively tested to reproduce well also the bandstructure of $sp^2$-hybridized AGNRs (see fig. \ref{fig:fig2}). Results show that the role of the buffer layer is not limited in the separation of the electronic properties of the substrate from that of graphene, but becomes an active component of the heterostructure's electronic behavior through energy states that are introduced from the dangling bonds of the SiC surface. These states pin the Fermi level of the system even in the case where the AGNR consists of up to three graphene layers, while such behavior is independent from the width of the considered nanoribbons. 

\begin{figure}
	\centering
		\includegraphics[width=0.6\columnwidth]{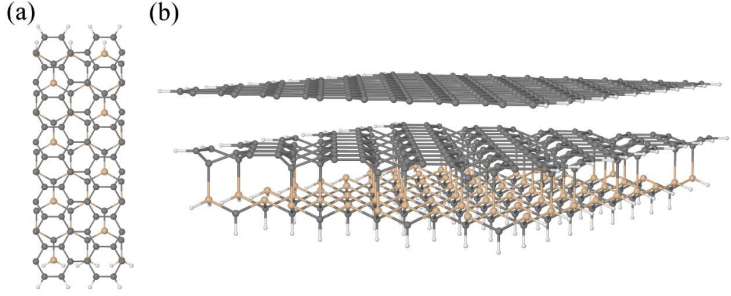}
	\caption{Geometrical representation of a 17 AGNR: (a) unit cell of the SiC(0001)/graphene interface with a $\sqrt3\times\sqrt3R30^{\circ}$ surface reconstruction (top view) and (b) single-layer AGNR (side view).}
	\label{fig:geom}
\end{figure}

We consider graphene nanoribbons on SiC substrates on the basis of relaxation information of ref. \cite{2007PhRvL..99g6802M} obtained by \textit{ab initio} molecular dynamics within the spin local density approximation. The surface here is reconstructed in a numerically more convenient  $\sqrt3\times\sqrt3R30^{\circ}$ basis that does not significantly alter the physics with respect to the experimental case. In this study, the substrate is formed by a single bilayer of SiC which is responsible for all bonding interactions with the carbon buffer layer, while hydrogen saturation has been imposed towards the bulk (fig. \ref{fig:geom}). The buffer layer stands at $d=2.58 $\AA{} over the SiC substrate, while interface atoms that covalently bond relax at $d=2$\AA{}. We will consider the composite substrate-buffer system as the base system from now on. Subsequent graphene layers follow graphite's ideal planar interlayer relaxation($d=3.35$\AA). Considering a unit cell $n$ of the 1D periodic structure, we calculate the bandstructure of various AGNRs by direct diagonalization of the $k$-space Hamiltonian matrix
\begin{equation}
 	[H(\vec{k})]=\sum_m {[H]_{nm}e^{i\vec{k} \cdot(\vec{d}_m - \vec{d}_n)}},
\label{eq:hamiltonian}
\end{equation}
where $\vec{k}$ is the Block wavevector within the first Brilouin zone. The summation over $m$ runs over all neighboring unit cells with which unit cell $n$ has any overlap (including itself) \cite{Datta}, while matrices $[H]_{nm}$ are written in real space on the previously discussed set of orbital functions\footnote{For $n \neq m$, $[H]_{nm}$ are interaction matrices between neighboring unit cells, whereas in the case of $n=m$, $[H]_{nn}$ refers to the Hamiltonian matrix of unit cell $n$.}. Vectors $\vec{d}_m$ and $\vec{d}_n$ show the position of two equivalent points of unit cells $m$ and $n$. The generalized eigenvalue equation reads \cite{2006JAP...100d3714K}: 
\begin{equation}
  [H(\vec{k})]\Psi_i(\vec{k})=E_i(\vec{k})[S(\vec{k})]\Psi_i(\vec{k}),
\label{eq:eigenvalue}
\end{equation}
where $[S(\vec{k})]$ is the $k$-space overlap matrix calculated in an analogous to eq. \ref{eq:hamiltonian} way, and $\Psi_i(\vec{k})$ ($E_i(\vec{k})$) is the eigenvector (eigenvalue) of the $i^{th}$ subband.

\begin{figure}
	\centering
		\includegraphics[width=0.6\columnwidth,angle=-90]{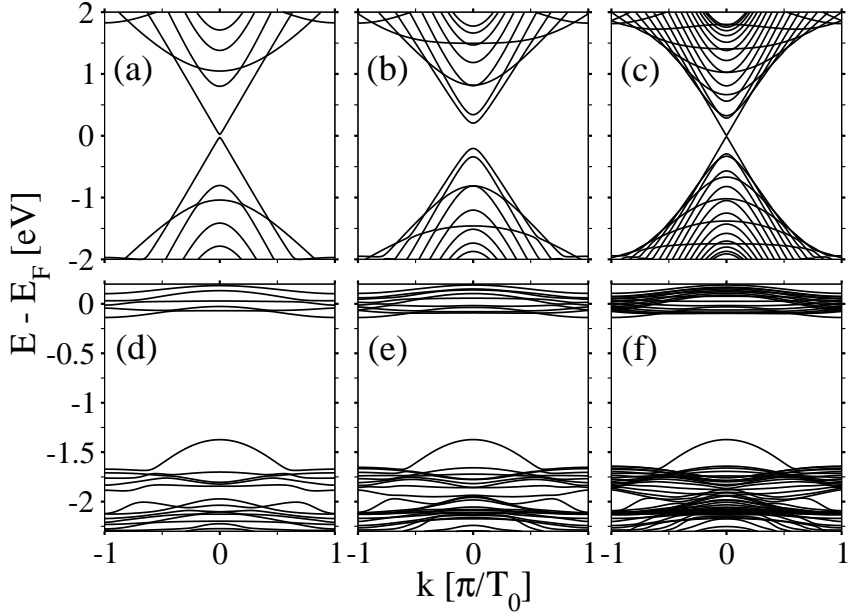}
	\caption{Bandstructure of an ideal (a)  11 AGNR, (b)  19 AGNR and (c) 35 AGNR. SiC(0001)/graphene interface systems corresponding to a (d) 11 AGNR, (e) 19 AGNR and (f) 35 AGNR. $T_0$ denotes the translation vector of the respective 1D periodic structure.}
	\label{fig:fig2}
\end{figure}

\begin{figure}
	\centering
		\includegraphics[width=0.6\columnwidth,angle=-90]{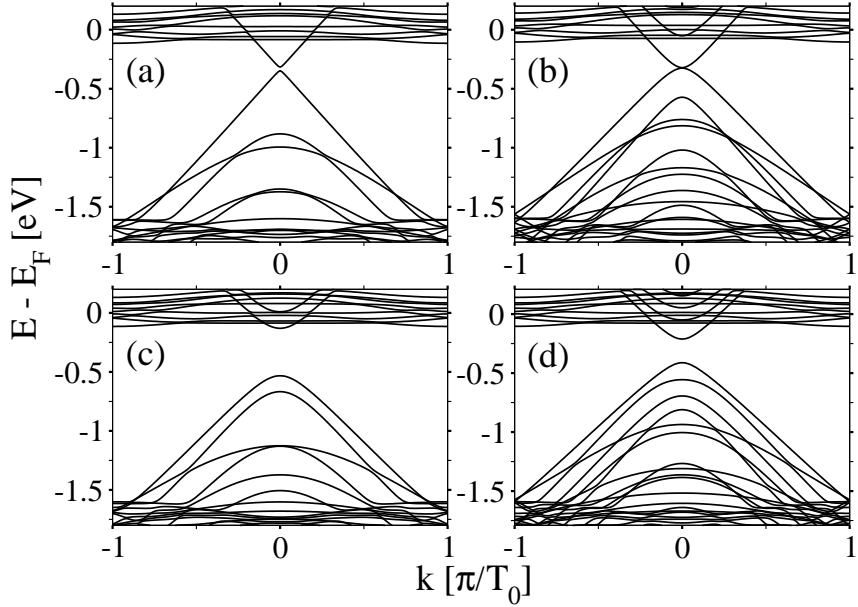}
	\caption{Bandstructure for AGNRs grown over the SiC(0001)/graphene interface system: (a) a single-layer 17 AGNR, (b) a double-layer 17 AGNR, (c) a single-layer 19 AGNR and (d) a double-layer 19 AGNR. $T_0$ denotes the translation vector of the respective 1D periodic structure.}
	\label{fig:fig3}
\end{figure}

We start off in an ideal reconstruction of the epitaxial process, from the SiC substrate up to 5-layer nanoribbons, considering various semimetallic and semiconducting AGNRs (consistent with the terminology of ref. \cite{son:216803}). The single-bilayer 1D substrate presents a dispersion relation with an indirect gap of $E_g=3.53eV$ for all cases, closer to the $\alpha$-SiC polytypes. Such deviation from the 3D bulk case can be strictly attributed to the perpendicular confinement and results in a shift of the conduction band towards higher energies. The first step towards graphitization is achieved by the formulation of the interface buffer layer. Here, the breaking of symmetry for the two complementary triangular sublatices that constitute the ideal graphene structure results in a complete alteration of the typical AGNR bandstructure, with the presence of a wide energy zone that for the most part is not crossed by any subband, while the Fermi level of the system is captured by weakly dispersive midgap bands (fig. \ref{fig:fig2}). It is interesting to note that due to deviation from planarity and selective covalent bonding, the buffer layer looses also the typical characteristic of 1D-confined graphene, that is the dependence of its electronic structure from its width. The process continuous with the successive addition of extra graphene layers on the systems that contrary to the buffer layer are Van der Waals bonded and preserve their $sp^2$ planar topology. The first layer upon the buffer one restores typical AGNR behavior \cite{son:216803}, namely an almost linear dispersion curve with a secondary confinement-induced bandgap for the semimetallic AGNRs, and the presence of wider gap in the semiconducting case (see fig. \ref{fig:fig3} for a 17 and a 19 AGNR). The fundamental concept thought is that the Fermi level remains pinned by the almost flat subbands of the base system, imposing and overall metallic character and $n$-type doping for all types of nanoribbons. This trend continuous with the addition of a second layer, with ribbons showing the typical parabolic dispersion curve below the trapped Fermi level. Results show that such behavior continues for up to three layers of confined graphene, while from the forth layer and onwards, the highest occupied subband gradually disperses towards graphite's Fermi level position for small values of the wavevector. This picture brings to discussion two important aspects: a) the main electronic properties of two-dimensional epitaxial graphene on SiC \cite{2007PhRvL..99g6802M} are confirmed also in the case of 1D graphene nanoribbons. b) The important role of confinement for suspended or $SiO_2$-deposited graphene is strongly compromised in the case of SiC substrates due to the Fermi level pinning effect and the intrinsic metallic behaviour of all AGNRs. 

We now move forward to better understand the importance of Fermi level positioning for the conductive characteristics of SiC grown AGNRs. For this purpose we follow a real-space approach with the objective of identifying the topological distribution of the density states for energies around the Fermi level. We use the non-equilibrium Green's function formalism for a 1D device with ideal semi-infinite contacts. The Green matrix reads:
\begin{equation}
[G]=(E[S] - [H] - [\Sigma_{left}] - [\Sigma_{right}])^{-1},
\label{eq:Green}
\end{equation}
where $E$ is the energy, $[H]$ ($[S]$) is the Hamiltonian (overlap) matrix written in real-space on the same basis as before, and the $[\Sigma_{left,right}]$ matrices introduce the role of scattering due to the left and right semi-infinite contacts. The latter are calculated within an optimized iterative scheme \cite{0305-4608-15-4-009}. The device spectral function is the anti-hermitian part of the Green matrix $[A(E)]=i([G]-[G^{\dagger}])$, from which the local density of states (LDOS) at energy $E$ and position $\vec{r_{\alpha}}$ can be defined as: 
\begin{equation}
 LDOS(\vec{r_{\alpha}},E) = \int_{-\infty}^{+\infty} \frac{Trace[A(E)][S]}{2 \pi} \delta(\vec{r}-\vec{r_{\alpha}}) d\vec{r} ,
 \label{LDOS}
\end{equation}
where $\delta$ is the Delta function and $\vec{r_{\alpha}}$ shows the positions of the atomic sites. Integrating eq. \ref{LDOS} over all the atomic positions results in the total density of states of the system. Finally, transport calculation can be introduced in the formalism by the definition of the zero-bias transmission coefficient $T=Trace([\Gamma_{left}][G][\Gamma_{right}][G^{\dagger}])$, where $[\Gamma_{left,right}]$ are the contact spectral functions \cite{Datta}.

\begin{figure}
	\centering
		\includegraphics[width=0.9\columnwidth]{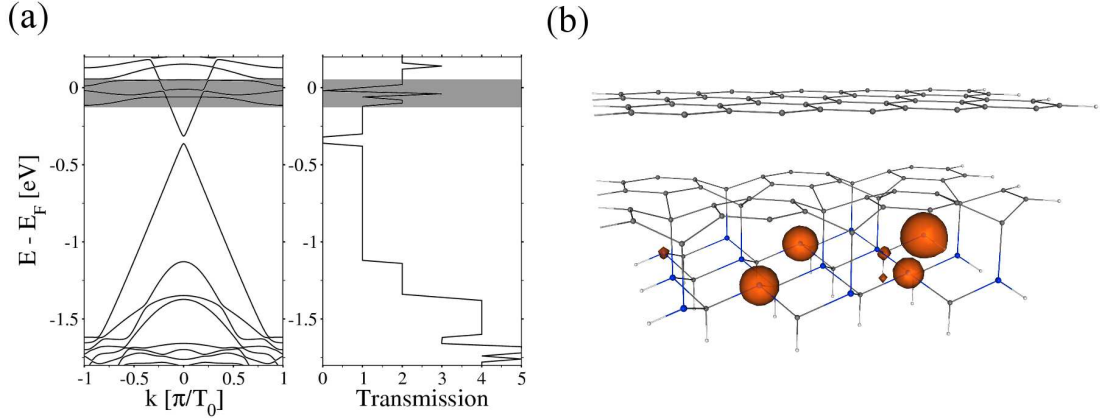}
	\caption{(a) Bandstructure and transmission as a function of energy for an 11 AGNR. (b) Schematic real-space representation of the LDOS corresponding to the grey energy region of figure \ref{fig:ldos}(a). The radius of each sphere is proportional to the amplitude of the LDOS value on that atomic site. $T_0$ denotes the translation vector of the respective 1D periodic structure.}
	\label{fig:ldos}
\end{figure}

We consider an 11 AGNR with a single graphene layer over the base system and calculate the LDOS for energies around the Fermi level position (fig. \ref{fig:ldos}). Results show that the almost flat subbands around the Fermi level are primarily projected in the $xyz$ space on the highly localized positions of the substrate's non-saturated Si-bonds. Such phenomenon raises a key issue for the conductive capacity of these systems, since it implies that the substrate becomes an active component of the conduction process, whereas the unsaturated Si atomic sites constitute conductive interface defects. Moreover such implication suggests that electrical conduction through epitaxial SiC nanoribbons can lead to a great reduction of the carrier mobility with respect to the $SiO_2$-deposited case and could be also related to the different conducting characteristics observed for epitaxial graphene on $(000\bar{1})$ and $(0001)$ substrates. It is worth mentioning that as soon as we leave the energy zone of the weakly dispersed interface subbands and move towards lower energies that restore the typical nanoribbon bandstructure, LDOS is located on the single graphene layer. It can be therefore argued that by perpendicular electric field modulation (e.g. using a gate electrode) the system's Fermi level can be detached from the substrate states and electrical conduction can recover the typical graphene characteristics. 

In conclusion, this letter presents bandstructure and local density of states calculations of armchair graphene nanoribbons epitaxially grown on SiC (0001). The main aspect of this study is the importance of the SiC(0001)/graphene interface dangling bonds, which introduce states that pin the Fermi level of the system even in the case of few-layer AGNRs. Such effect can have an adverse impact on the conductive capacity of these systems since it creates an electron transport channel through interface defects, compromising in an non-trivial way device-related properties like high-carrier mobility. This feature also brings to attention the role of confinement, since in the absence of gate modulation, structures are always expected to behave in a metallic way. Finally, the present study implies that a thorough understanding of surface reconstruction is necessary for the conductive properties, and hence, for determining the device application of graphene on SiC substrates. 

\begin{acknowledgments}
The authors would like to acknowledge the MIUR for partial financial support.
\end{acknowledgments}

\end{document}